\documentclass[12pt,preprint]{aastex}

\usepackage{graphicx}
\usepackage{amsmath}

\def \apj {ApJ}

\def \apjl {ApJ}
\def \solphys {Solar Phys.}

\def \aap {A\&A}

\newcommand{\citeN}[1]{\citeauthor{#1} (\citeyear{#1})}
\newcommand{\citeNP}[1]{\citeauthor{#1} \citeyear{#1}}

\newcommand{\mxcm}{Mx~cm$^{-2}$}

\shortauthors{Socas-Navarro & Lites}
\shorttitle{Small-Scale Mixture of Field Strengths in the Quiet Sun}

\begin{document}

\title{Observational Evidence for Small-Scale Mixture of Weak and Strong
  Fields in the Quiet Sun}

\author{H. Socas-Navarro}
   	\affil{High Altitude Observatory, NCAR\thanks{The National Center
	for Atmospheric Research (NCAR) is sponsored by the National Science
	Foundation.}, 3450 Mitchell Lane, Boulder, CO 80307-3000, USA}
	\email{navarro@ucar.edu}

\author{B.W. Lites}
   	\affil{High Altitude Observatory, NCAR, 3450 Mitchell Lane, Boulder,
   	CO 80307-3000, USA} 
	\email{lites@ucar.edu}


\begin{abstract}
Three different maps of the quiet Sun, observed with the Advanced Stokes
Polarimeter (ASP) and the Diffraction-Limited Stokes Polarimeter (DLSP), show
evidence 
of strong ($\simeq$1700~G) and weak ($<$500~G) fields coexisting within the
resolution element both at network and internetwork locations. The angular
resolution of the observations is of 1\arcsec 
(ASP) and 0.\arcsec6 (DLSP), respectively. Even at the higher DLSP
resolution, a significant fraction of the network magnetic patches harbor a
mixture of strong and weak fields. Internetwork elements that exhibit kG
fields when analyzed with a single-component atmosphere are also shown to
harbor considerable amounts of weak fields. Only those patches for which
a single-component analysis yields weak fields do not show this mixture of
field strengths. Finally, there is a larger fractional area of weak fields in
the convective upflows than in the downflows.
\end{abstract}
   
\keywords{line: profiles --  Sun: atmosphere -- Sun: magnetic fields --
                Sun: photosphere}

\section{Introduction}
\label{sec:intro}

Our understanding of quiet Sun magnetic fields is evolving at a very
rapid pace. The classical picture of the quiet Sun is based on a
sprinkling of kilo-Gauss (kG)
fluxtube-like structures of sub-arcsecond sizes (unresolved in the
observations) forming a magnetic network at the
boundaries of the 
supergranular cells. The internetwork (cell interiors) would be almost devoid
of flux, exhibiting only sparse weak turbulent flux concentrations. As new
observations became available 
with improved sensitivity and spatial resolution, the
internetwork region started to gain importance. Strong kG fields were also
found in many internetwork locations and the amount of magnetic flux and
energy detected has increased as the instrumentation improved
(\citeNP{KDE+94}; \citeNP{L95}; \citeNP{GDKS96}; \citeNP{LR99};
\citeNP{SAL00}; \citeNP{L02}; 
\citeNP{SNSA02}; \citeNP{KCS+03}; \citeNP{DCKSA03}; \citeNP{DCSAK03};
\citeNP{SNMPL04}; \citeNP{LSN04}).

At the same time, the concept of organized magnetic structures (e.g.,
fluxtubes) as the building blocks for quiet Sun fields seems to be losing
some ground, at least outside of the network. Recent numerical simulations
(\citeNP{C99}; \citeNP{EC01}; \citeNP{SAEC03}; \citeNP{S03}; \citeNP{SN02})
reveal a more disorganized, 
almost chaotic, scenario with the field being dragged around by turbulent
convective motions. Another important element in this picture is the
discrepancy in the distribution of internetwork fields as seen in visible and
infrared observations. Infrared data (\citeNP{L95}; \citeNP{KCS+03}) show a
predominance of sub-kG fields, with a distribution that peaks around 350~G. On
the other hand, authors working with visible observations (e.g.,
\citeNP{SNSA02}) obtain that most of these fields are of kG
strength. \citeN{SNSA03} proposed that this discrepancy is a natural
consequence of unresolved small-scale inhomogeneities of the field. If one
has a mixture of weak and strong fields coexisting in the resolution element,
visible and infrared observations tend to emphasize different parts of the 
distribution. 
  
In a recent paper, \citeN{SN04b} (see also \citeNP{SN04a}) showed that the
visible \ion{Fe}{1} lines at 6302 \AA \, exhibit some sensitivity to
unresolved field strength inhomogeneities. In the present work we make use of
this property to seek evidence that mixed strengths indeed occur in
the quiet Sun. As we discuss below, we are able to detect a rather large
number of such mixed strengths. Our results support the ``disorganized''
picture of the quiet Sun fields discussed above, not only in the internetwork
but also to some extent in the magnetic network.

\section{Observations and analysis}
\label{sec:obs}

The datasets analyzed in this paper come from two different instruments. The
Advanced Stokes Polarimeter (ASP) is a spectro-polarimeter for the Dunn Solar
Telescope (DST) at the Sacramento Peak observatory (Sunspot, NM, USA),
operated by 
the National Solar Observatory. We used the map observed by \citeN{L96} on
Sep 29, 1994 (hereafter referred to as Map~1). This map has a spatial
resolution of $\simeq$1\arcsec \, and a spectral resolution of
$\simeq$30~m\AA . The spectral lines observed are the pair of \ion{Fe}{1} lines
in the 6302 \AA \, region. 

The other instrument employed is the Diffraction-Limited Stokes Polarimeter
(DLSP), which is operated at the same telescope. The DLSP is a new
instrument that has been designed specifically to take advantage of the new
adaptive optics (AO) system at the DST in order to achieve very high angular
resolution. It has been optimized for the routine observation
of the 6302 \AA \, spectral region. 

\citeN{LSN04} obtained what can
be considered very high-resolution Stokes observations of the quiet Sun,
of approximately 0.\arcsec6. This figure, alongside with the 0.\arcsec5
reached by \citeN{DCKSA03}, represent the highest resolution
spectro-polarimetric observations of quiet Sun fields made thus far. In our
analysis here we consider the best two maps observed by \citeN{LSN04}, namely
the ones observed on Sep 14 and Sep 16 2003 (hereafter Map~2 and Map~3,
respectively). 


\subsection{One-component analysis}
\label{sec:1c}

We first conducted a one-component (hereafter 1C) analysis, assuming that we
have  one field strength ($B$) that occupies a certain area filling factor
($\alpha$) of the spatial pixel. Instead of performing an iterative
least-squares fit to the observations, we 
chose to do forward modeling from a large number of models. In this manner we
make sure that the entire model space is probed and that the
absolute minimum of the $\chi^2$ merit function is found. We start with two
models for the thermodynamic parameters (temperature, gas
pressure, micro- and macroturbulence and line-of-sight velocity) of the quiet
Sun, representing a granule and an intergranular lane. The models were
obtained from the inversion of average Stokes~I profiles.
The Stokes~V profiles were then synthesized for many values of the
magnetic field strength (ranging from $B=300$ to $B=2000$~G) and a global
velocity offset (from $v=-2.5$ to $v=2.5$ km~s$^{-1}$). The
calculations were performed with the code LILIA (\citeNP{SN01a}), assuming
LTE and hydrostatic equilibrium. 

\begin{figure}
\plotone{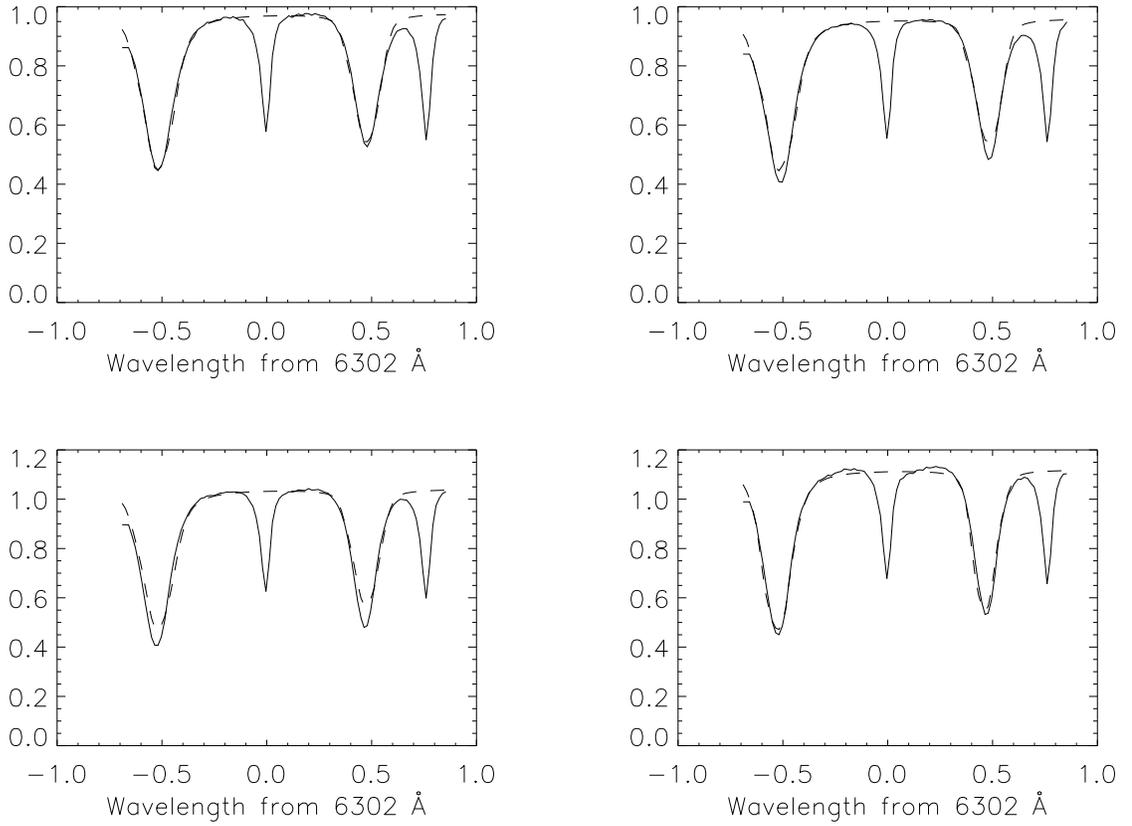}
\caption{
\label{intfits}
Fits (dashed) to observed (solid) intensity profiles corresponding
to (x,y) coordinates (20,60),(50,90),(80,120) and (110,150) of Map~2.
}
\end{figure}

Depending on the continuum intensity of the observations, we used the granule
or lane models or a suitable linear combination of both. In the case of the
quiet Sun this is approach is valid because the thermodynamical properties
of the atmosphere experience only relatively small variations across the
map. Fig~\ref{intfits} shows the fits obtained with this method to intensity
profiles from four randomly-chosen locations. Reasonably good fits are
obtained for the entire dataset analyzed, which justifies the approximation
used for the thermodynamics of the atmosphere.

The synthetic Stokes~V profiles ($V^{syn}$) 
were multiplied by a filling factor $\alpha=A(V^{obs})/A(V^{syn})$ (where $A$
denotes the amplitude of a given profile) and then compared one by one to the
observed profile ($V^{obs}$) at each spatial position of the maps. The sign
of the synthetic profile is chosen to match the polarity of $V^{obs}$.
The values of $B$, $v$ 
and $\alpha$ that yield the best fit to $V^{obs}$ are taken as representative
of the conditions at the spatial location under consideration. This process
is repeated for every pixel in every map. The spatial distributions of $\alpha$
and $B$ thus obtained are shown if Figs~\ref{fig:map1} and~\ref{fig:map23}
(upper panels). 

It is important to note that the 6302 \AA \, lines are in the weak field
regime (i.e., the regime in which the Zeeman splitting is much smaller than
the Doppler width) for field strengths lower than $\simeq$500~G. In the
weak field regime the shape of the Stokes~V profiles is independent of the
magnetic field. The profiles are simply scaled with the value of the
field. Thus, it is not possible to disentangle the effects of
the filling factor from the field strength. In other words, any value of the
field weaker than 500~G will result in exactly the same $\chi^2$. 
Therefore, the reader must 
keep in mind that any fields below 500~G depicted in the figures might be
actually weaker (with correspondingly larger filling factors).

\subsection{Two-component analysis}
\label{sec:2c}

\citeN{SN04b} suggested that the \ion{Fe}{1} lines at 6302 \AA 
\, exhibit some sensitivity to the presence of two magnetic strength
components. When the two field strengths are at least as far apart as 500~G
and 1700~G, their response functions are sufficiently decoupled that the
{\it relative} filling factors of these two components can be inferred with
an uncertainty of 0.10 or less. 

As a further step in our study, we carried out a two-component (2C) analysis
of the observations. 
Considering the arbitrariness in the strength of the weak component we
chose a value of 300~G, which is close to the peak of the distribution
obtained from 1C inversions of infrared internetwork observations
(\citeNP{KCS+03}). Thus, we (arbitrarily) fixed the strengths of the weak and
strong components to 300~G and 1700~G, respectively. The problem now is to
find the filling factors $\alpha_w$ and $\alpha_s$. Let us define these filling
factors relative to the magnetic element, so that $\alpha_w +
\alpha_s = 1$. The total magnetic filling factor in the observed pixel is
still $\alpha$. The filling factors of the two components relative to the
resolution element are then $\alpha \alpha_w$ and $\alpha \alpha_s$. This
convention may seem somewhat confusing at first, but it is useful for the
discussion in \S\ref{sec:results} below.

The synthetic profiles $V^{syn}$ for the 2C case were calculated in the
following manner. As before, we start with the a model for the atmospheric 
thermodynamics which depends on the observed continuum intensity. We then
synthesized the profiles $V^w$ and $V^s$ for 300 
and 1700~G, respectively. At each spatial location we compared the profiles
$V^{obs}$ and $\alpha V^{syn}=\alpha (\alpha_w V^w + \alpha_s V^s)$ (where,
again, $\alpha=A(V^{obs})/A(V^{syn})$). The combinations of
$\alpha$, $\alpha_w$, $\alpha_s$ and $v$ that
lead to the best fit of the observations are 
selected. Notice that only two of these
parameters, $v$ and either $\alpha_s$ or $\alpha_w$, are independent.

In order to ensure that the detection of mixed field strengths 
has significance, we rewarded solutions with either $\alpha_w=0$ or
$\alpha_s=0$. The $\chi^2$ corresponding to these solutions is reduced by
10\%. This conservative
approach makes the procedure ``prefer'' one-component solutions. It is also
important to point out that, while we are not adding any degrees of freedom
when going from 1C (above) to 2C, the $\chi^2$ is smaller in
the 2C analysis virtually everywhere. These arguments (alongside with those
in \S\ref{sec:reliability}) give us confidence in
the results reported below.

\section{Results}
\label{sec:results}

\begin{figure*}
\vspace{-0.5in}
\epsscale{0.9}
\plotone{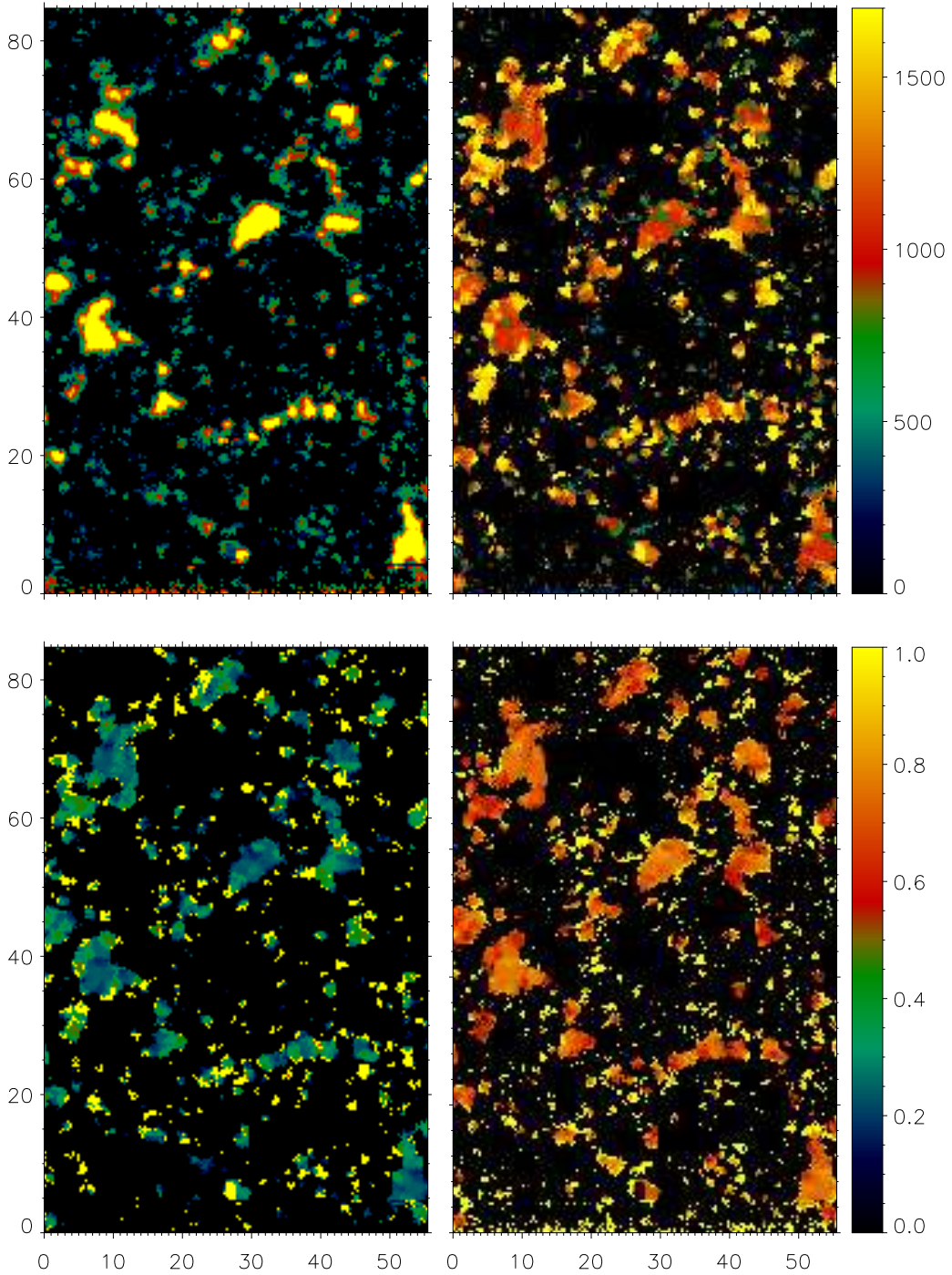}
\caption{Map 1 (ASP). Upper left: Total magnetic filling factor, saturated at
  0.10. The color scale in this panel ranges from 0 to 0.10. Upper right: Field
  strength inferred from the 1C inversion (G). Lower 
  left: Relative filling factor of the strong field component. Lower right:
  Relative filling factor of the weak field component. In the lower left
  (right) image the color scale represents the percentage of the magnetic
  area occupied by strong (weak) fields. Spatial units are
  arc-seconds. Pixels appearing black in both 
  images do not exhibit polarization signal above the noise and have been
  excluded from the analysis. 
\label{fig:map1}
}
\end{figure*}

\begin{figure*}
\plotone{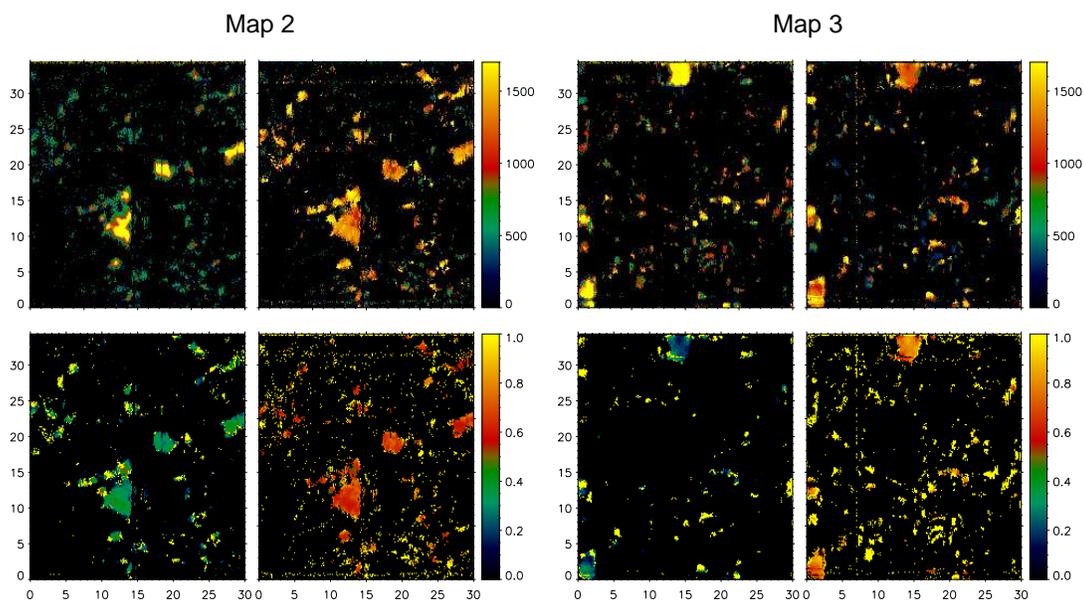}
\caption{Maps 2 and 3 (DLSP). Four panels represent the following quantities
  for each map. Upper left: Total magnetic filling factor,
  saturated at 
  0.10. The color scale in this panel ranges from 0 to 0.10. Upper right:
  Field strength inferred from the 1C inversion (G). Lower 
  left: Relative filling factor of the strong field component. Lower right:
  Relative filling factor of the weak field component. In the lower left
  (right) image the color scale represents the percentage of the magnetic
  area occupied by strong (weak) fields. Spatial units are
  arc-seconds. Pixels appearing black in both 
  images do not exhibit polarization signal above the noise and have been
  excluded from the analysis.
\label{fig:map23}
}
\end{figure*}

Figs~\ref{fig:map1} and~\ref{fig:map23} show the spatial distribution of
$\alpha$ (upper left), $B$ (upper right), $\alpha_s$ (lower left) and
$\alpha_w$ (lower right). The 1C analysis results in strong kG fields in most
spatial pixels inverted (i.e., those exhibiting significant polarization
signal), with the exceptions of very few weak field patches in the 
network and some weak field elements in the internetwork. This is consistent
with previous observations published in the literature based on visible
lines, in which the distribution of fields peaks around 1.5~kG (see
references in \S\ref{sec:intro}). 

The lower panels of the figures clearly show the presence of mixed-strength
pixels both in network and internetwork locations. Let us start by discussing
the results for the internetwork. In this region, visible and infrared
observations (always using 1C analyses) have led to disparate conclusions,
with the infrared lines showing a much larger fraction of weaker
fields. According to \citeN{SNSA03}, this can be explained by a small-scale
mixture of weak and strong fields beyond the spatial resolution. They showed
that, when such mixture exists, a 1C inversion of the visible lines is biased
towards the stronger fields. Therefore, we would expect to have mixed field
strengths in those spatial pixels where the 1C analysis results in strong
fields. Those with weak fields, on the other hand, are probably rather
homogeneous. This is exactly what we find in the analysis of our internetwork
profiles, as seen in Figs~\ref{fig:map1} and~\ref{fig:map23}, as well as in
Table~\ref{table:maps}. We did not find any mixed strengths in pixels with
$\alpha < 0.1$ and $B < 500$~G. 

Let us now turn to the network. The relatively strong flux concentrations in
network patches have been traditionally associated with strong kG fields (see
references in \S\ref{sec:intro}). While our 1C analysis agrees with this
assessment, the 2C analysis reveals the presence of a significant amount of
weak fields mixed at small scales. The mixed strengths occur all over the
network patches and not only 
around their perimeter. Table~\ref{table:maps} lists some properties of the
network elements, including the percentage of pixels showing mixed
strengths. Finally, we find (not shown in the table) that the fractional area
occupied by weak fields is larger in the convective upflows than in the
downflows (as one would expect).

\clearpage

\clearpage

\begin{deluxetable}{lccc}
  \tablewidth{0pt}
  \tablecaption{Flux and mixed-strengths statistics
     \label{table:maps}}
  \tablehead{            & Map 1 & Map 2 & Map 3 }
  \startdata
  Mean 1C flux  (\mxcm) &  8.7  & 5.2  & 6.2  \\
  Mean 2C flux  (\mxcm) &  9.2  & 5.4  & 6.1  \\
  Mean 1C field (G)     & 1529  &  999 & 1064 \\
  Mean 2C field (G)     & 1192  &  928 & 924  \\
  Pixels analyzed       & 25.9\% & 24.5\% & 24.0\% \\
  Pixels harboring      &       &      &      \\
   mixed strengths      & 14.8\% & 7.6\% & 2.8\% \\
  Mixed strength in pixels  &       &      &      \\
  with $\alpha < 0.1$, $B < 500$~G & 0\% & 0\% & 0\% \\
  Mean flux in          &        &       &       \\
   strong fields (\mxcm) & 6.3   &  3.3  & 2.6  \\
  Mean flux in          &        &       &       \\
   weak fields (\mxcm)  & 2.9    &  2.1  & 3.5   \\
  Fraction of upflowing         &       &        &        \\
   area with strong fields       &  12.9\% & 1.6\%    &  9.5\%  \\
  Fraction of downflowing        &       &            &      \\
   area with strong fields       & 28.2\%  &  24.4\%  &  13.2\%  \\
  \enddata
\end{deluxetable}

\clearpage

\section{Reliability of the results}
\label{sec:reliability}

We have not been concerned thus far with problems such as line asymmetries or
details of the sub-pixel field distribution. This would be a concern if we
were interested in obtaining a detailed description of the solar atmosphere,
including velocity and magnetic field gradients. Our goal, however, is not
that, but rather to
demonstrate that weak and strong fields generally coexist at sub-arcsecond
spatial scales. How is this overall conclusion affected by the simplifying
assumptions employed in our work? This important question is explored in the
present section by means of numerical simulations. Before going into the
details, though, it is probably a good idea to consider the broader context
of the subject in order to put our work into perspective.

In doing analysis of solar magnetic fields based on spectro-polarimetric
observations there 
are various levels of complexity that one can consider, depending on the
sophistication of the physical models employed. The most simplistic analyses
are the ``traditional'' techniques, such as the line ratio, the
separation of Stokes~V extrema, or fits by Gaussian profiles. These techniques
do not involve radiative transfer calculations and do not require a detailed
knowledge of the atmospheric model. They typically provide only a rough
approximation to the magnetic field, but are unable to deal with line
asymmetries due to unresolved inhomogeneities, atmospheric gradients, etc. 
The most sophisticated techniques
make use of full radiative transfer calculations in detailed model
atmospheres to fit the observed profiles in a least-squares sense (inversion
codes). 

Most of the published work (see references in
\S\ref{sec:intro}) makes use of the simplistic techniques to infer ``typical''
values of the photospheric field in the quiet Sun. These simplified analyses
led to the inference that network fields are typically strong
($\simeq$1.5~kG), or that internetwork fields appear to be mostly kG when
observed in the visible and sub-kG ($\simeq$300~G) when observed in the
infrared. Very few authors (e.g., the works of \citeNP{SAL00};
\citeNP{SNSA02}) have used inversions that consider detailed models with
atmospheric gradients. The overall picture (i.e., whether the fields are weak
or strong) does not change signifincantly, although the inversion codes
obviously provide 
more information. Detailed inversions have not been carried out in the
infrared so far. The analyses published in the literature are derived from
fits of Gaussian functions to the observed profiles.

Our analysis in this paper lies somewhere between the two
ends of the range of sophistication. We calculate actual Stokes profiles from
a model atmosphere, but do not consider line-of-sight variations of the
quantities. The thermodynamics of our models is probably
reasonably good, as suggested by the fits shown in
Fig~\ref{intfits}. However, there are two important simplifications. First,
we do not consider line-of-sight gradients of the field. Second, we
parameterize the unresolved field distribution by two discrete field values
of 300~G and 1700~G. It is important to note that none of these assumptions
are worse than those of the traditional methods (save for the exceptions
noted above) which are based on single-valued fields without height
variations. 

We have carried out some numerical simulations in order to gain insights
into the adequateness of our approximations. First consider how our two-point
approach is representative of the sub-pixel field distribution.
Suppose that we have an unresolved probability
distribution function (PDF) of field strengths $f(B)$. The filling factor of
fields with a strength between $B$ and $B+\Delta B$ is:

\begin{equation}
\alpha(B,B+\Delta B)=\int_B^{B+\Delta B} f(B) dB \, .
\end{equation}

The emerging profile $P(\lambda)$ from such distribution is:

\begin{equation}
P(\lambda)=\int_0^\infty f(B) P_B(\lambda) dB \, ,
\end{equation}
where $P_B(\lambda)$ denotes the profile produced by a field of strength
$B$. For the moment we are only interested in horizontal
inhomogeneities. The effects of gradients of $B$ along the line of sight will
be discussed later. If we consider $P(\lambda)$ as a simulated observation
and apply to it the procedure introduced in \S\ref{sec:2c} above, we can
compare the inferred values with the actual PDF, $f(B)$,
employed. Table~\ref{table:pdf} shows the results of various tests with 
different shapes of the PDF. As proposed by \citeN{EC01},
and by \citeN{KCS+03}, \citeN{SNSA03}, and \citeN{TBSAR04}, we adopted an
exponential dependence\footnote{
It should be noted that the PDF obtained by \citeN{EC01} from the
simulations is actually somewhat more complicated. Their PDF is a
stretched exponential in the absence of net unsigned flux, and has a
shoulder when such flux is non-zero. 
} for $f(B)$ with a
normalization factor $N^{-1}=\int_0^\infty f(B) dB$. The filling factor of
fields stronger than 2~kG has been set to zero. We also considered a
Dirac-delta PDF ($\delta(B-B_0)$), which simply means that the field is
homogeneous over the resolution element, with a strength $B_0$. 

\begin{deluxetable}{ccccc}
  \tablewidth{0pt}
  \tablecaption{Results from simulations. Inferrences using values of 300~G
  and 1700~G. 
     \label{table:pdf}}
  \tablehead{ PDF   & Actual & Inferred & Actual & Inferred \\
              ($B$ in G) & $\alpha(B<$1~kG) & 
    $\alpha(B=300$~G) & $\alpha(B>$1~kG) &
               $\alpha(B=1.7$~kG) }
  \startdata
  $f=\delta(B-100)$ & 1.00 & 1.00 & 0.00 & 0.00 \\
  $f=\delta(B-2000)$ & 0.00 & 0.00 & 1.00 & 1.00 \\
  $f=N\exp(-B/100)$ & 1.00 & 1.00 & 0.00 & 0.00 \\
  $f=N\exp(-B/300)$ & 0.96 & 0.93 & 0.04 & 0.07 \\
  $f=N\exp(-B/600)$ & 0.84 & 0.82 & 0.16 & 0.18 \\
  $f=N\exp(-B/1000)$ & 0.73 & 0.75 & 0.27 & 0.25 \\
  $f=N\exp(-B/1500)$ & 0.66 & 0.70 & 0.34 & 0.30 \\
  $f=N\exp(-B/2000)$ & 0.62 & 0.67 & 0.38 & 0.33 \\
  \enddata
\end{deluxetable}

The results in Table~\ref{table:pdf} suggest that the filling factors that we
obtained for $B=300$~G and $B=1700$~G are more or less representative of the
weak and strong fields present in more complex distributions expected to be
present in the actual quiet Sun. We also tested the sensitivity of our
results to the values that we used to represent weak and
strong fields. To this aim we repeated the calculations above but this time
using 
500 and 1500~G. The results obtained, listed in Table~\ref{table:pdf2}, are
very similar to those of Table~\ref{table:pdf}. 

\begin{deluxetable}{ccccc}
  \tablewidth{0pt}
  \tablecaption{Results from simulations. Inferrences using values of 500~G
  and 1500~G. 
     \label{table:pdf2}}
  \tablehead{ PDF   & Actual & Inferred & Actual & 
              Inferred \\
              ($B$ in G) & $\alpha(B<$1~kG) & 
    $\alpha(B=500$~G) & $\alpha(B>$1~kG) &
              $\alpha(B=1.5$~kG) } 
  \startdata
  $f=\delta(B-100)$ & 1.00 & 1.00 & 0.00 & 0.00 \\
  $f=\delta(B-2000)$ & 0.00 & 0.00 & 1.00 & 1.00 \\
  $f=N\exp(-B/100)$ & 1.00 & 1.00 & 0.00 & 0.00 \\
  $f=N\exp(-B/300)$ & 0.96 & 0.97 & 0.04 & 0.03 \\
  $f=N\exp(-B/600)$ & 0.84 & 0.79 & 0.16 & 0.21 \\
  $f=N\exp(-B/1000)$ & 0.73 & 0.77 & 0.27 & 0.33 \\
  $f=N\exp(-B/1500)$ & 0.66 & 0.60 & 0.34 & 0.40 \\
  $f=N\exp(-B/2000)$ & 0.62 & 0.57 & 0.38 & 0.43 \\
  \enddata
\end{deluxetable}

Let us now consider the issue of velocity and magnetic field gradients along
the line of sight. Such gradients give rise to asymmetries in the line
profiles. We introduced gradients in our simulated observations and tested
their effects on the filling factors inferred by our analysis. The
line-of-sight velocity that we introduced varies linearly 
between $\tau_{500}=10$ and $\tau_{500}=10^{-3}$ (with $\tau_{500}$ denoting
the optical depth at 500~nm). The amplitude of the
variation is of 2~km~s$^{-1}$. The field strength varies linearly as well,
from $B$ at $\tau_{500}=10^{-1}$ (with $B$ ranging from 0 to 2~kG) to zero at
$\tau_{500}=10^{-3}$. The asymmetric profiles calculated from the PDFs above
are again taken as simulated observations and applied our analysis (which
neglects gradients). The results from this experiment are summarized in
Table~\ref{table:pdf3} (the numbers within parentheses are obtained when 
the velocity gradient is doubled). These values indicate that the presence of
asymmetries does not invalidate our conclusions on the coexistence of field
strengths in the resolution element.

\begin{deluxetable}{ccccc}
  \tablewidth{0pt}
  \tablecaption{Results from simulations with asymmetric profiles.
     \label{table:pdf3}}
  \tablehead{ PDF   & Actual & Inferred & Actual & 
              Inferred \\
              ($B$ in G) & $\alpha(B<$1~kG) & 
    $\alpha(B=500$~G) & $\alpha(B>$1~kG) &
              $\alpha(B=1.5$~kG) } 
  \startdata
  $f=\delta(B-100)$ & 1.00 & 1.00 (0.97) & 0.00 & 0.00 (0.03) \\
  $f=\delta(B-2000)$ & 0.00 & 0.01 (0.00) & 1.00 & 0.99 (1.00) \\
  $f=N\exp(-B/100)$ & 1.00 & 0.99 (0.95) & 0.00 & 0.01 (0.05) \\
  $f=N\exp(-B/300)$ & 0.96 & 0.91 (0.87) & 0.04 & 0.09 (0.13) \\
  $f=N\exp(-B/600)$ & 0.84 & 0.80 (0.75) & 0.16 & 0.20 (0.25) \\
  $f=N\exp(-B/1000)$ & 0.73 & 0.73 (0.75) & 0.27 & 0.27 (0.25) \\
  $f=N\exp(-B/1500)$ & 0.66 & 0.69 (0.63) & 0.34 & 0.31 (0.37) \\
  $f=N\exp(-B/2000)$ & 0.62 & 0.66 (0.60) & 0.38 & 0.34 (0.40) \\
  \enddata
\end{deluxetable}

\section{Conclusions}
\label{sec:conc}

This paper reports on observational evidence for mixed
field strengths in the quiet Sun ($\sim$300 and $\sim$1700~G) on spatial
scales smaller than 0.\arcsec6. Mixed strengths are found in network and
internetwork magnetic elements. The results for the internetwork are not
entirely unexpected. \citeN{SNSA03} proposed that such mixture is the
most natural explanation for the discrepancy between visible and
infrared observations. Our work strongly supports their
conjecture and starts to bridge the gap between the two types of
observations. The presence of weak fields inside network patches is more
surprising and had not been anticipated before.

The reliability of our results is backed by several facts. Our 1C analysis
reproduces what had been obtained in the past from visible
spectro-polarimetric observations. When we extend the analysis to 2C we find 
that the merit function $\chi^2$ is systematically smaller, even though we
did not add free parameters in the model. Therefore, our 2C analysis is at
least as reliable as comparable 1C studies which have been used thus
far. In order to ensure a conservative criterion for the occurrence of 2C,
our procedure was implemented with a ``preference'' for 1C solutions when
possible. 
The spatial distributions of the filling factors $\alpha_w$ and $\alpha_s$ are
smooth, exhibiting spatial coherence even though each pixel has been inverted
separately. These 
distributions, as well as the other results obtained from our study, are
consistent in the three maps analyzed (which have been obtained from
two different instruments). Several numerical tests (\S\ref{sec:reliability})
show that our method is able to distinguish between the filling
factors of weak and strong fields in the presence of unresolved PDFs and/or
line asymmetries. Finally, our results are sensible from a physical
point of view. For example, we find mixed strengths in internetwork locations
showing strong 1C fields but not in those showing weak 1C fields. Moreover,
the relative fractional area occupied by weak and strong fields is different
for upflows and downflows, with the downflows having a larger filling factor
of kG fields (in agreement with existing simulations).

It is important to keep in mind that the mixed strengths that we have
detected may be just the tip of the iceberg. It is very likely that we only
see the most conspicuous ones. We have chosen to parameterize the sub-pixel
field distribution by two discrete values at 300 and 1700~G. However, we 
know from both observations and simulations that quiet Sun fields obey a
continuous probability distribution, with the weaker fields covering a larger 
fraction of the resolution element. Unfortunately, according to
\citeN{SN04b}, we can only infer (at most) two points of such distribution
using visible observations. Simultaneous visible and infrared observations
will offer a much more detailed picture. It might then be possible to infer
more than two points of the distribution.


\end{document}